\documentclass[11pt]{article}
\pdfoutput=1
\usepackage{amssymb}
\usepackage{graphicx}
\newcommand\cl[1]{\centerline{#1}}
\newcommand{\fig}[2]{\includegraphics[width=#1\textwidth]{#2}}

\newcommand {\apj}      {The Astrophysical Journal}
\newcommand {\apjl}     {The Astrophysical Journal Letters}

\newcommand{\Non}{\ensuremath{N_\mathrm{on}}}
\newcommand{\Nbg}{\ensuremath{N_\mathrm{bg}}}

\begin{document}

\title{Are EeV cosmic rays isotropic at intermediate scales?}
\author{M. Yu.\ Zotov\\
Skobeltsyn Institute of Nuclear Physics,\\
Lomonosov Moscow State University, Russia
}
\date{}
\maketitle

\begin{abstract}

	We study anisotropy of cosmic rays in the energy range 0.2--1.4~EeV at
	intermediate angular scales using the public data set of the Pierre Auger
	Observatory. At certain scales, the analysis reveals a number of deviations
	from the isotropic distribution with the statistical significance up to~4
	standard deviations.
	It also demonstrates that the anisotropy evolves with energy.
	If confirmed with the complete Auger or Telescope Array data sets,
	the result can shed
	new light on the structure of galactic magnetic fields and the problem of
	transition from galactic to extragalactic cosmic rays.

\end{abstract}

\section{Introduction}

Studies of anisotropy of cosmic rays (CRs) with energies around 1~EeV
($10^{18}$~eV) have a long and fascinating history. 
The first results date back to early 1960's~\cite{Linsley_etal-isotropy-1962}
but an intensive work is still in progress in both
theoretical and experimental directions.
In recent years, the Pierre Auger Observatory and the Telescope Array project
performed a great amount of sophisticated studies of anisotropy of ultra-high
energy CRs, and some of them were dedicated to EeV energies.
The studies were mostly performed in two directions:
(i)~large-scale anisotropy, such as an analysis of
dipole and quadrupole anisotropy, and
(ii)~small-scale anisotropy, namely searches for point sources 
of neutrons or photons (regions with radius $\lesssim2^\circ$).
No significant deviation from isotropy was revealed at large scales within the
systematic uncertainties, and no point sources were found,
see~\cite{Auger-Deligny-10PeV-2014,Deligny-2014-ankle,TA-summary-2013}
and references therein.

Some studies have also been performed at intermediate angular scales.
In 2005, a region around the Galactic Center (GC) was analyzed
in the energy range 1.0--2.5~EeV with smoothing at an angular scale of
$13.3^\circ$ and at energies 0.8--3.2~EeV at 
$3.7^\circ$ (and a small scale of $1.5^\circ$)~\cite{Auger-galcenter-2005}.
A $\pm10^\circ$ band around the Galactic plane (GP) was studied in the energy
range 1--5~EeV with smoothing on a~$10^\circ$ scale.
A blind search for localized excess fluxes in the whole field of view (FoV) of
the Auger experiment was performed in two energy bands above 1~EeV (1--5~EeV
and $\ge5$~EeV) at angular scales of $5^\circ$ and
$15^\circ$~\cite{Auger-localized-2005}.
The study employed 29073 events in the lower energy band.

In 2007, the anisotropy studies around the GC were updated with increased
statistics~\cite{Auger-galcenter-2007}.
Anisotropy of arrival directions of events in the energy range 0.97--3.16~EeV
was studied using circular windows of~$5.5^\circ$.
The direction to the GC was also studied at energies in the ranges
0.1--1~EeV and 1--10~EeV for windows sizes of $10^\circ$ and
$20^\circ$~\cite{Santos-galcenter-2007}.

The latest results of a blind search for localized excesses in four energy
ranges above 1~EeV
in circular windows of angular radius of $5^\circ$ and $15^\circ$ 
over the full exposed sky were presented in~\cite{Auger-Revenu-2013}.
All these studies gave results compatible with isotropy.
The only exception was a result of 2011~\cite{Lyberis-thesis-2011}, when
a region around the GC was studied with circular cells of radii
extending from~$2^\circ$ to~$20^\circ$ in the energy range 0.6--3.8~EeV, and
an excess was found at an~$8^\circ$ scale for energies above 0.9~EeV.
The excess was only observable in winter months though and was
concluded to be due to seasonal effects.

The Telescope Array collaboration studied anisotropy of 
events with energies 1--2.5~EeV and 0.7--1.8~EeV
registered with the surface detector~\cite{TA-summary-2013,TA-medium-2013}.
The event density map was averaged over the circles of $20^\circ$ radius
centered on the $1^\circ\times1^\circ$ grid.
No significant deviations from isotropy were found.
Table~\ref{tab:summary} gives a summary of the parameters of these searches.
To the best of the author's knowledge, the KASCADE-Grande experiment, which
was able to register cosmic rays up to 1~EeV, has never published results
concerning anisotropy at intermediate scales.

\begin{table}[!ht]
	\caption{Summary of studies of anisotropy of CRs registered with
		Auger and Telescope Array around 1~EeV at intermediate angular scales.}
	\label{tab:summary}
	\begin{center}\begin{tabular}{|c|c|c|c|}
		\hline
		Energy range, EeV	& Radius of circular windows	& Field	& Ref. \\
		\hline
		0.8--3.2	&	$3.7^\circ$					&	GC			&	\cite{Auger-galcenter-2005} \\
		1.0--2.5	&	$13.3^\circ$				&	GC			&	\\
		1.0--5.0	&	$10^\circ$					&	GP			&	\\
		\hline
		1.0--5.0	&	$5^\circ$, $15^\circ$	&	Auger FoV&	\cite{Auger-localized-2005} \\
		\hline
		0.97--3.16&	$5.5^\circ$					&	GC			&	\cite{Auger-galcenter-2007}\\
		\hline
		0.1--1.0	&	$10^\circ$, $20^\circ$	&	GC			&	\cite{Santos-galcenter-2007}\\
		\hline
		0.6--3.8	&	$2^\circ$--$20^\circ$	&	GC			&	\cite{Lyberis-thesis-2011} \\
		\hline
		1.0--2.0	&	$5^\circ$, $15^\circ$	&	Auger FoV&	\cite{Auger-Revenu-2013} \\
		\hline
		1.0--2.5	&	$20^\circ$					&	TA FoV	&	\cite{TA-summary-2013,TA-medium-2013}\\
		\hline
	\end{tabular}\end{center}
\end{table}

Thus, an interval of energies just below 1~EeV has not been
studied at intermediate scales in the full field of view of the recent
experiments yet.
Such an analysis seems to be interesting since conclusive information on
anisotropy in this interval can shed light on one of the fundamental problems
of astrophysics, namely the energy of transition from galactic to
extragalactic cosmic rays, see~\cite{Deligny-2014-ankle}, as well as on the
structure of the galactic magnetic field.
Besides this, the discovery of localized regions of excess of cosmic rays in
the TeV--PeV energy range, where the CR flux is more isotropic at large scales
than around 1~EeV, demonstrated that the distribution of arrival directions of
cosmic rays can have unexpected patterns at certain intermediate scales even
though they are isotropic at large scales, see~\cite{DiSciascio-Iuppa-review-2014}
for a recent review.
It was argued in~\cite{Kachelriess_etal-2007} that
under certain conditions the galactic magnetic field can induce
anisotropies in the observed flux of extragalactic CRs in models that
invoke a dominating extragalactic proton component already at $E\simeq0.4$~EeV.
This short work has two main goals: to try to figure out if there are any
statistically significant deviations from isotropy at intermediate angular
scales around 1~EeV, including an interval below 1~EeV,
and if so, to check if anisotropy evolves with energy.

\section{The Data and The Analysis Technique}

To tackle the problem, we used the public data set of the Pierre Auger
Observatory\footnote{\texttt{http://auger.colostate.edu/ED}}.
In order to deal with as many events as possible, we have chosen
the energy interval from 0.2 to 1.4~EeV.
As of June~7, 2014, it contained 30474 events.
In what follows, we shall call this ``the main data set.''
To study the evolution of anisotropy with energy, we also analyzed two subsets
of the main data set: one from 0.2~EeV to 0.56~EeV and another from 0.56~EeV to
1.4~EeV with 0.56~EeV being the median value of energies of events in the main
set.
The lower energy (LE) set consists of 15184 events, the higher energy
(HE) one consists of 15290 events.

The well-known shuffling (time swapping) technique was used for the analysis.
The method was developed independently by the Fly's
Eye~\cite{FlysEye-CygX3-1989} and CYGNUS~\cite{Alexandreas_etal-1991}
experiments and used later with minor modifications by multiple collaborations,
including Telescope Array and Auger, see,
e.g.,~\cite{TA-summary-2013,Auger-galcenter-2005,Auger-galcenter-2007,Lyberis-thesis-2011}.
It allows one to obtain a background map of arrival directions
of events registered by a particular experiment under the assumption of their
isotropy.
The method employs multiple cycles of swapping arrival times and
arrival directions of registered events.
An averaged map is taken then as an estimate of the background flux.
In our case, we divided the field of view of the Pierre Auger Observatory in
$0.2^\circ\times0.2^\circ$ bins and performed 200,000 cycles of shuffling.
The maximum difference between consecutive background maps at the end of
the procedure was less than $2\cdot10^{-5}$ for the main data set and of
the order of $1.5\cdot10^{-5}$ for the LE and HE subsets with
the relative difference $<0.5$\%.

We also performed calculations using a simulated isotropic background.
To do this, we generated one million data sets with the same number of events
as in the main data set and the same distribution w.r.t.\ declination
but uniformly distributed w.r.t.\ right ascension.\footnote{%
	The directional exposure of the Auger experiment in right ascension
	is known to be slightly non-uniform because of the tilt of the 
	array~\cite{Auger-deAlmeida-2013} but this is ignored here.}
A map of such isotropic arrival directions was calculated for each of the
data sets, and the average was taken as an estimate of the isotropic background
(IBG).
The IBG map was compared to the real map the same way as
the background map obtained with the shuffling technique.

To calculate the statistical significance~$S$ of deviations from the background, we
used the so called Li--Ma significance (formula~(17) from~\cite{LiMa-1983}),
which is a standard in anisotropy studies by Auger, Telescope Array and other CR
experiments.
One has to be accurate when choosing regions for the analysis with this formula
because of a small amount of data available.
Simple simulations show that~$S$ gives a satisfactory approximation to the
Gaussian distribution if one considers regions with the number of background
events $\Nbg\gtrsim200$ for the main data set and $\Nbg\gtrsim100$ for the
subsets.
One should also take into account that~$S$ slightly underestimates the
statistical significance of deviations if~$\Nbg$ is of the order of
a few percent of the whole number of events.

\section{The Main Results}

\begin{figure}[!ht]
	\cl{\fig{.42}{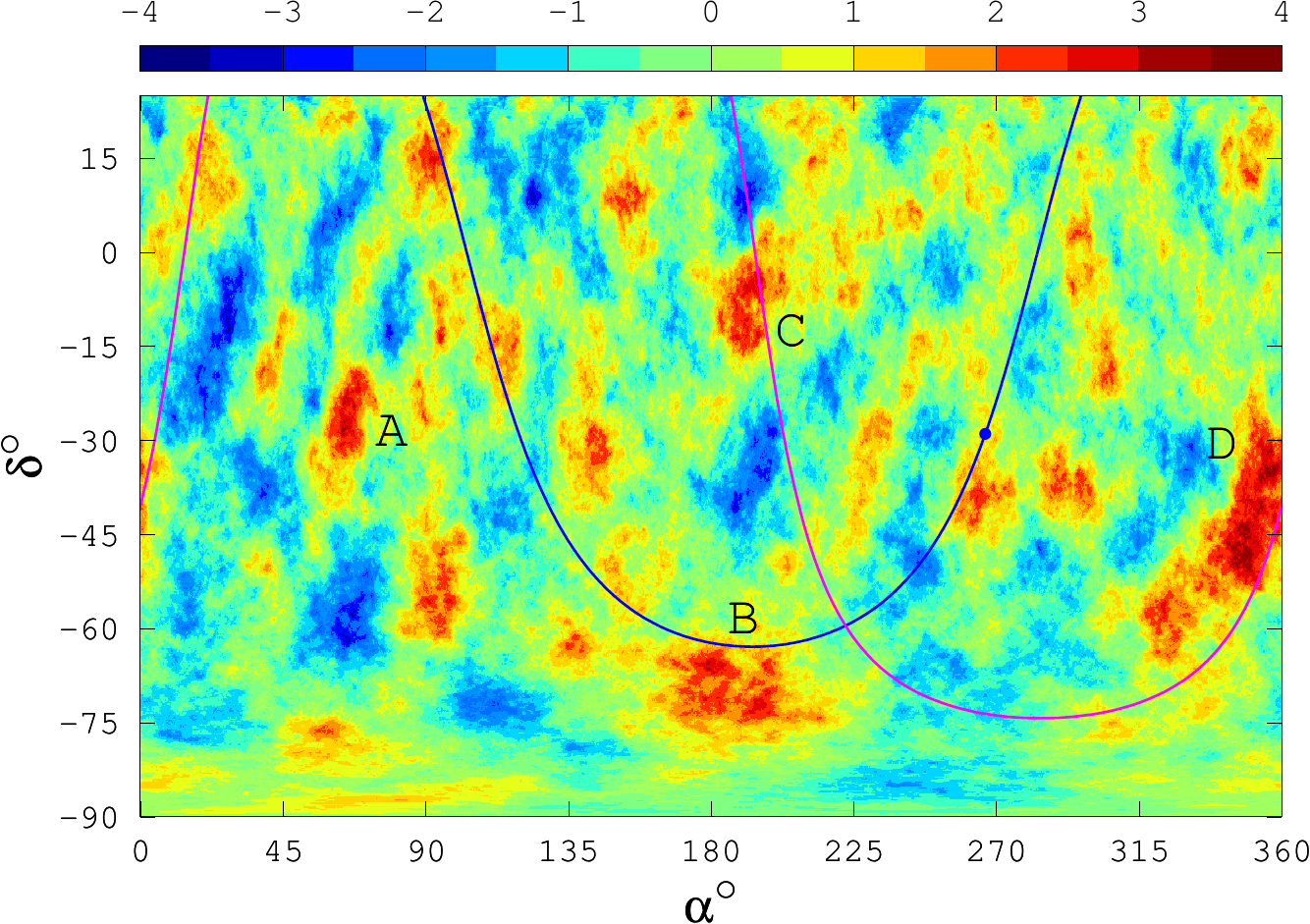}	\fig{.42}{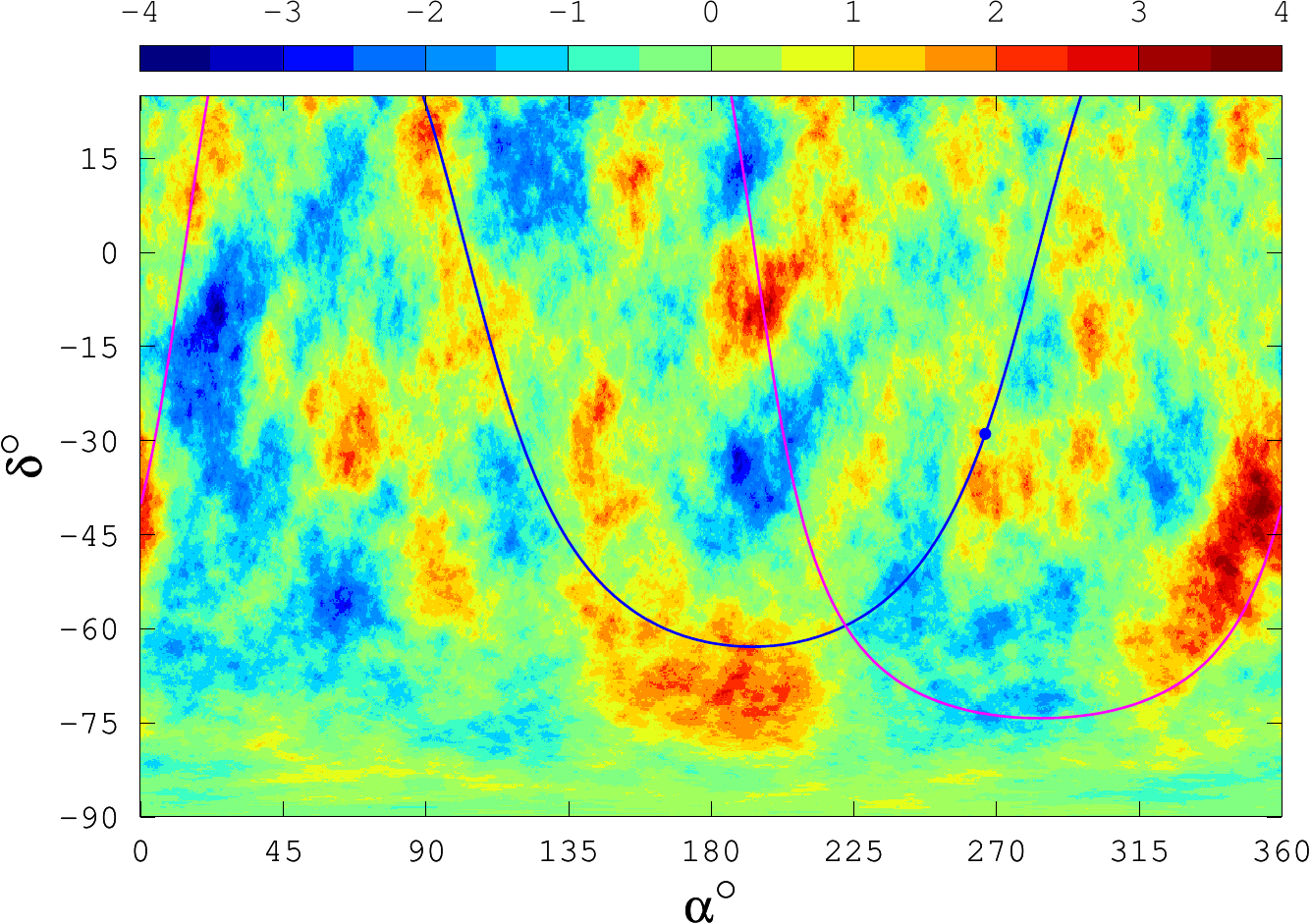}}
	\cl{\fig{.42}{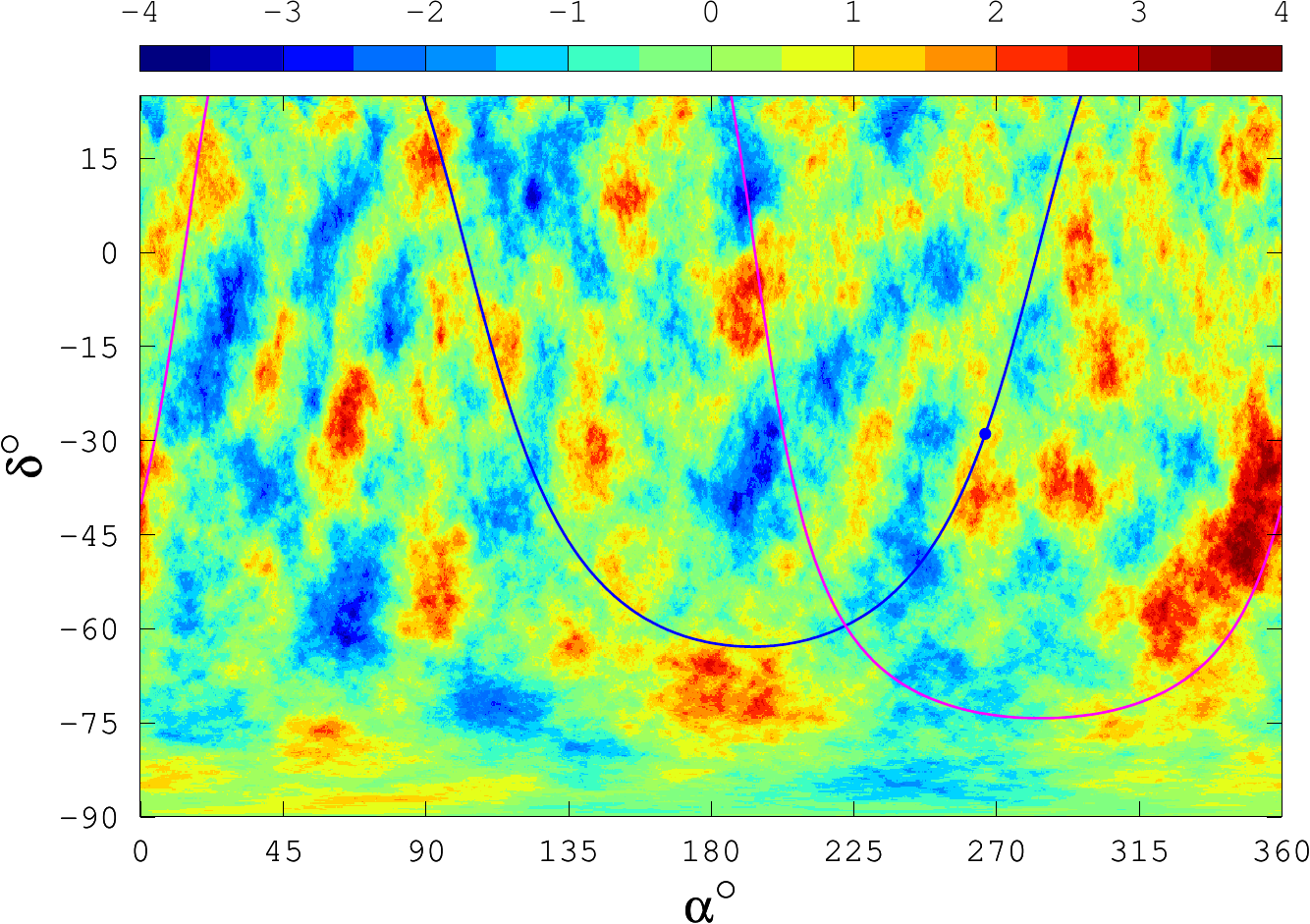}	\fig{.42}{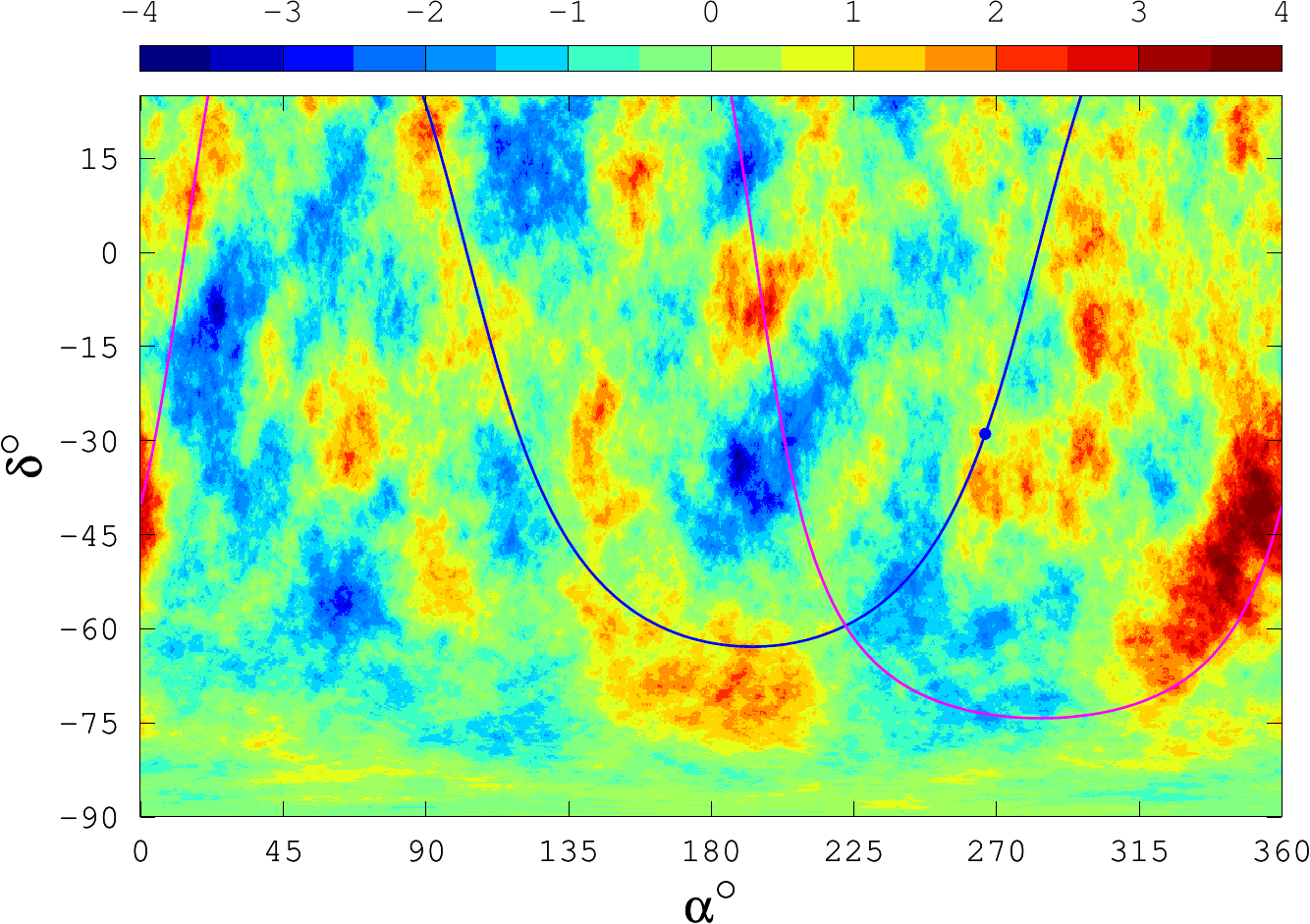}}
	\cl{\fig{.42}{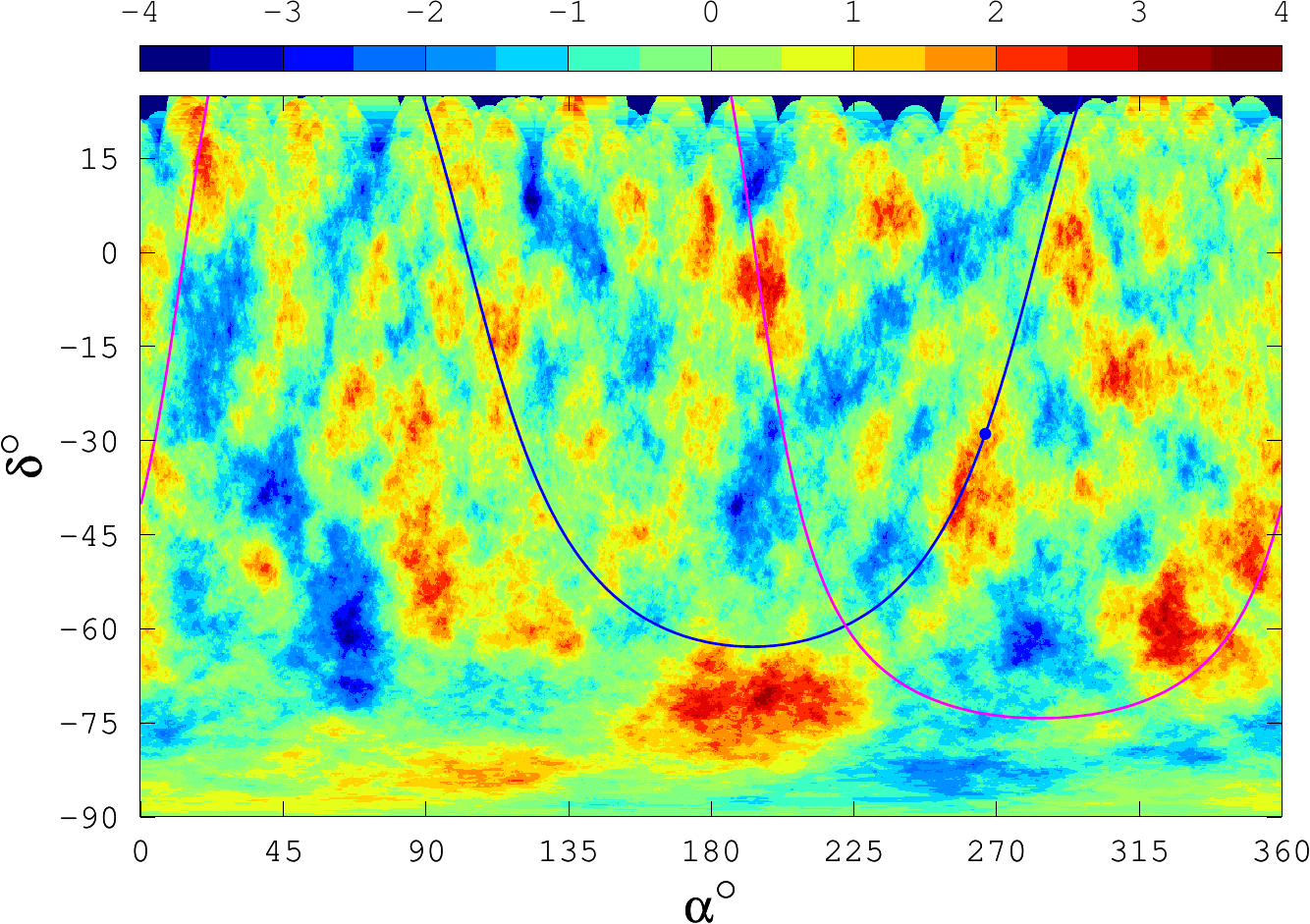}		\fig{.42}{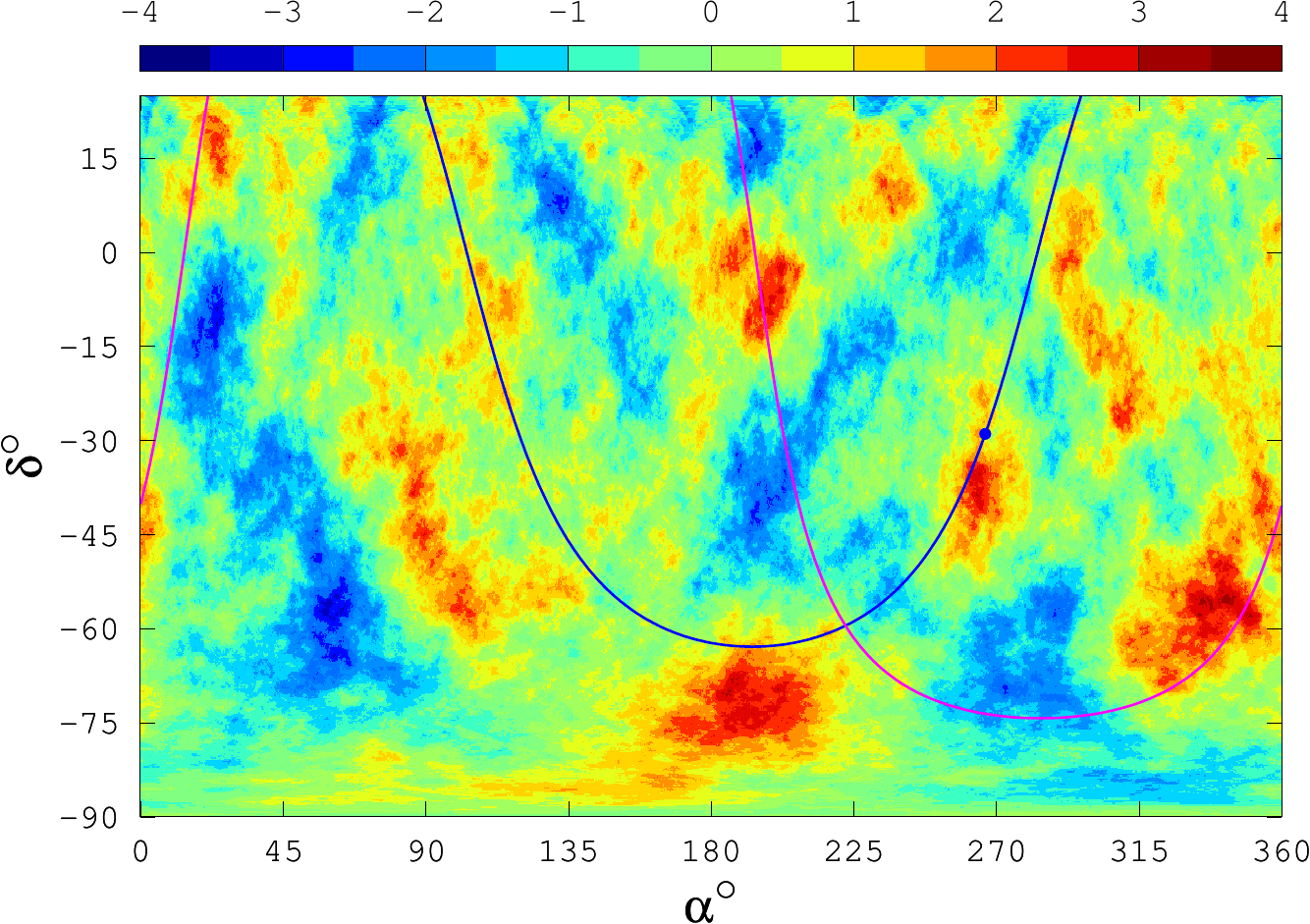}}
	\cl{\fig{.42}{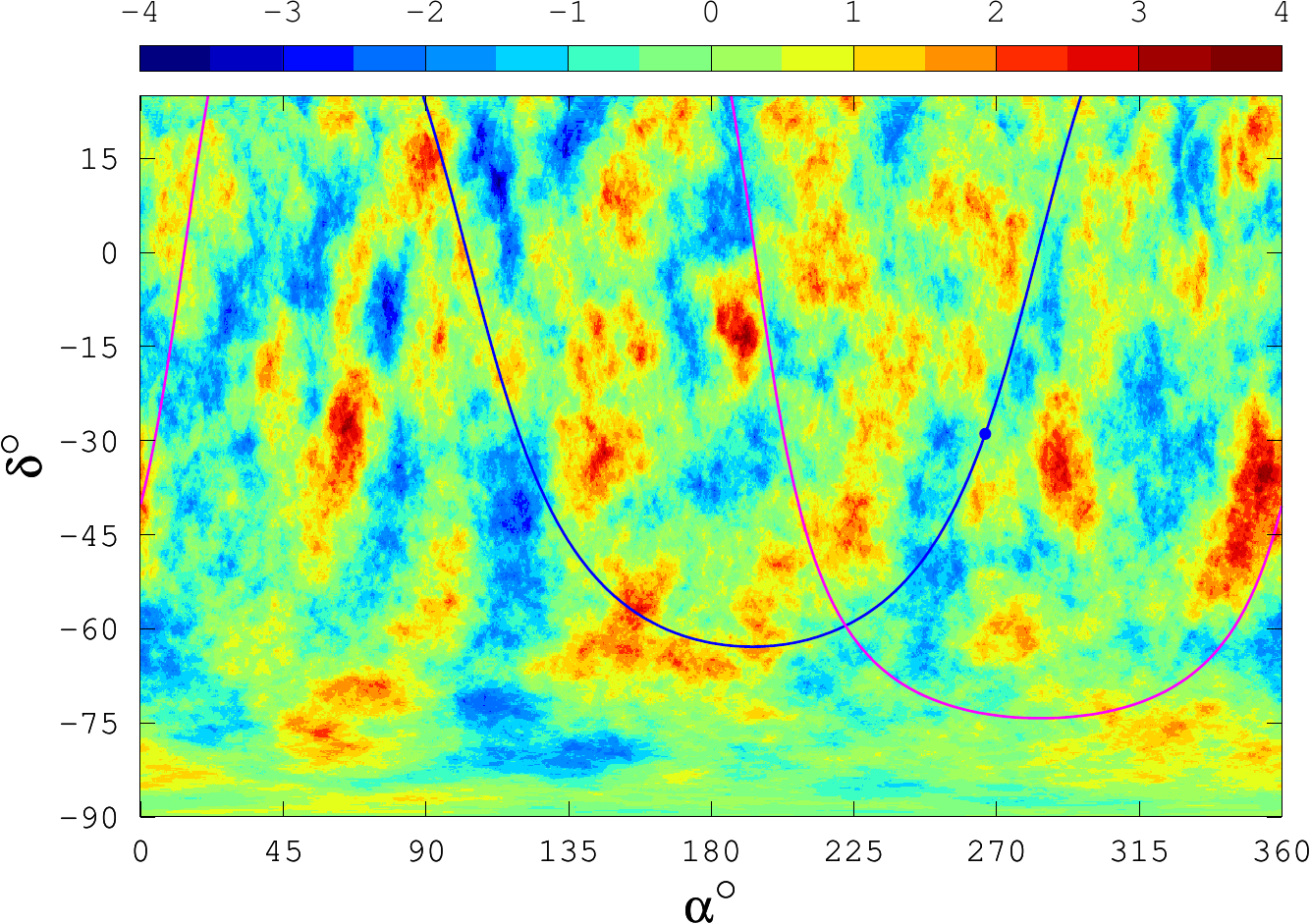}		\fig{.42}{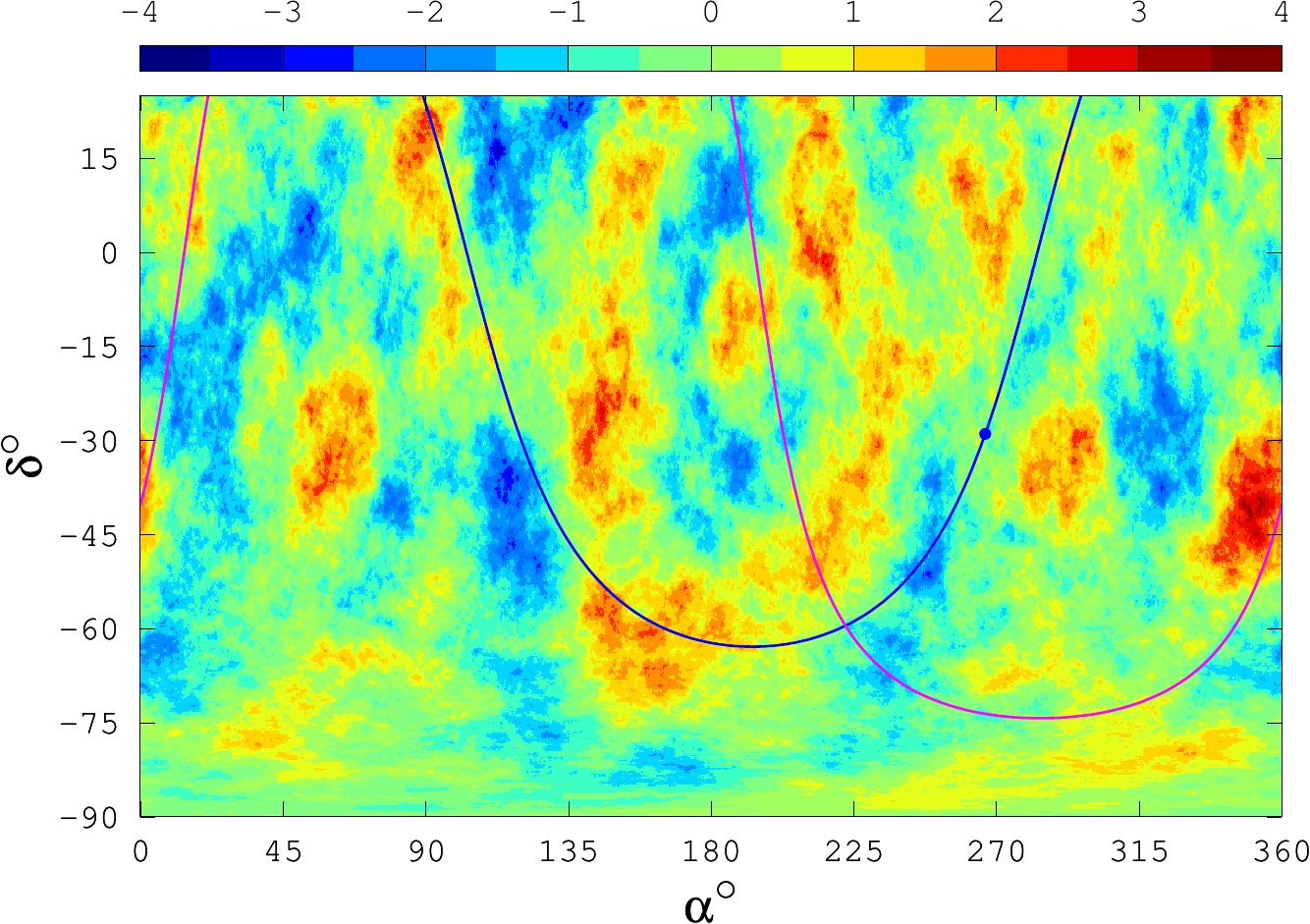}}
\caption{%
	From top to bottom: maps of anisotropy for the main data set,
	the main data set on the IBG, the LE and HE data sets.
	The maps are smoothed at $R=8^\circ$ (left column) and $R=12^\circ$ (right column).
	Equatorial coordinates are used. The blue curve shows the Galactic plane with
	the Galactic Center indicated by a bold point. The magenta curves show
	the Supergalactic plane.
	Colors indicate the Li--Ma significance of deviation from the expected
	background.
}
\label{fig:R8R12}
\end{figure}

\begin{figure}[!ht]
	\cl{\fig{.42}{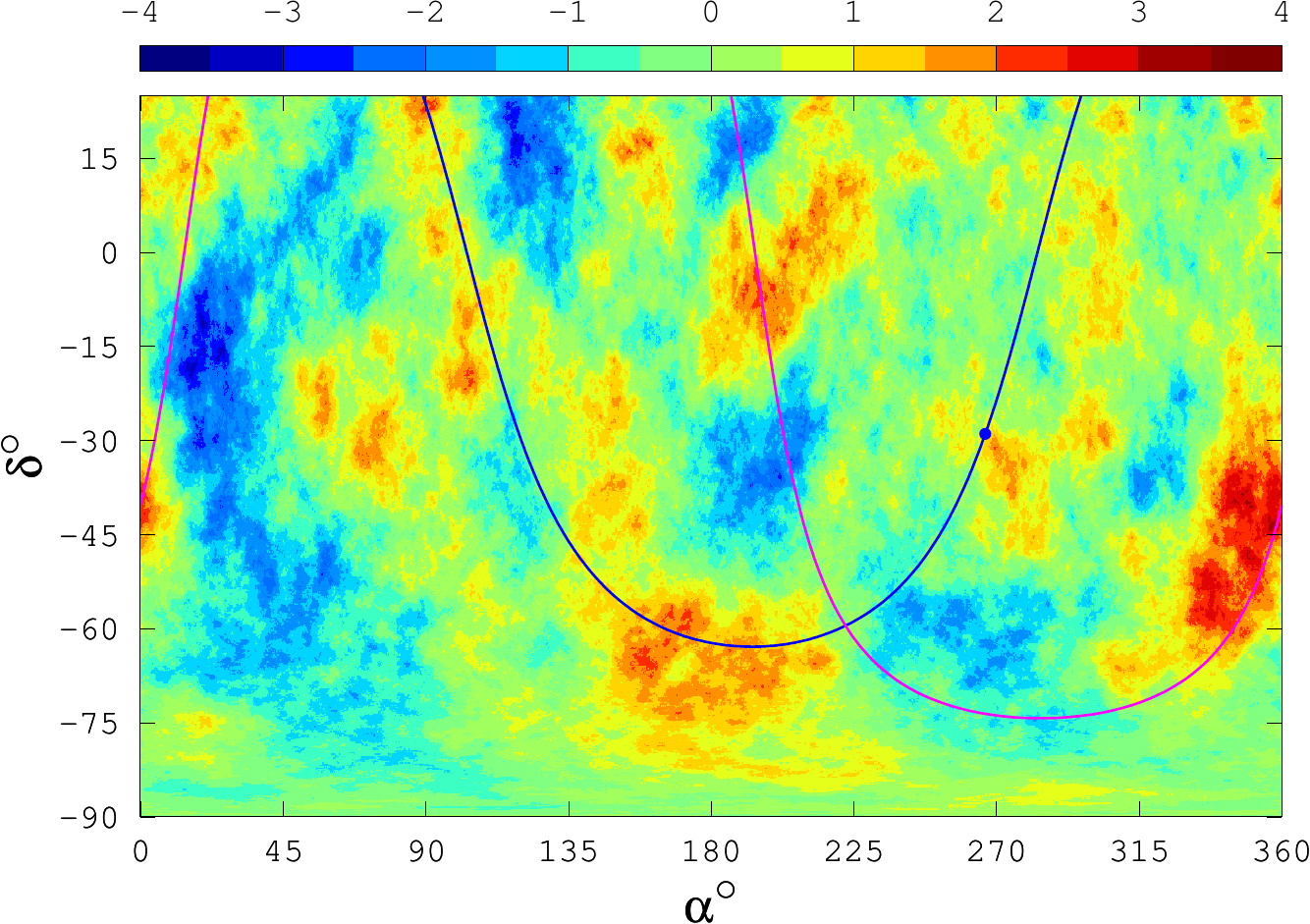}		\fig{.42}{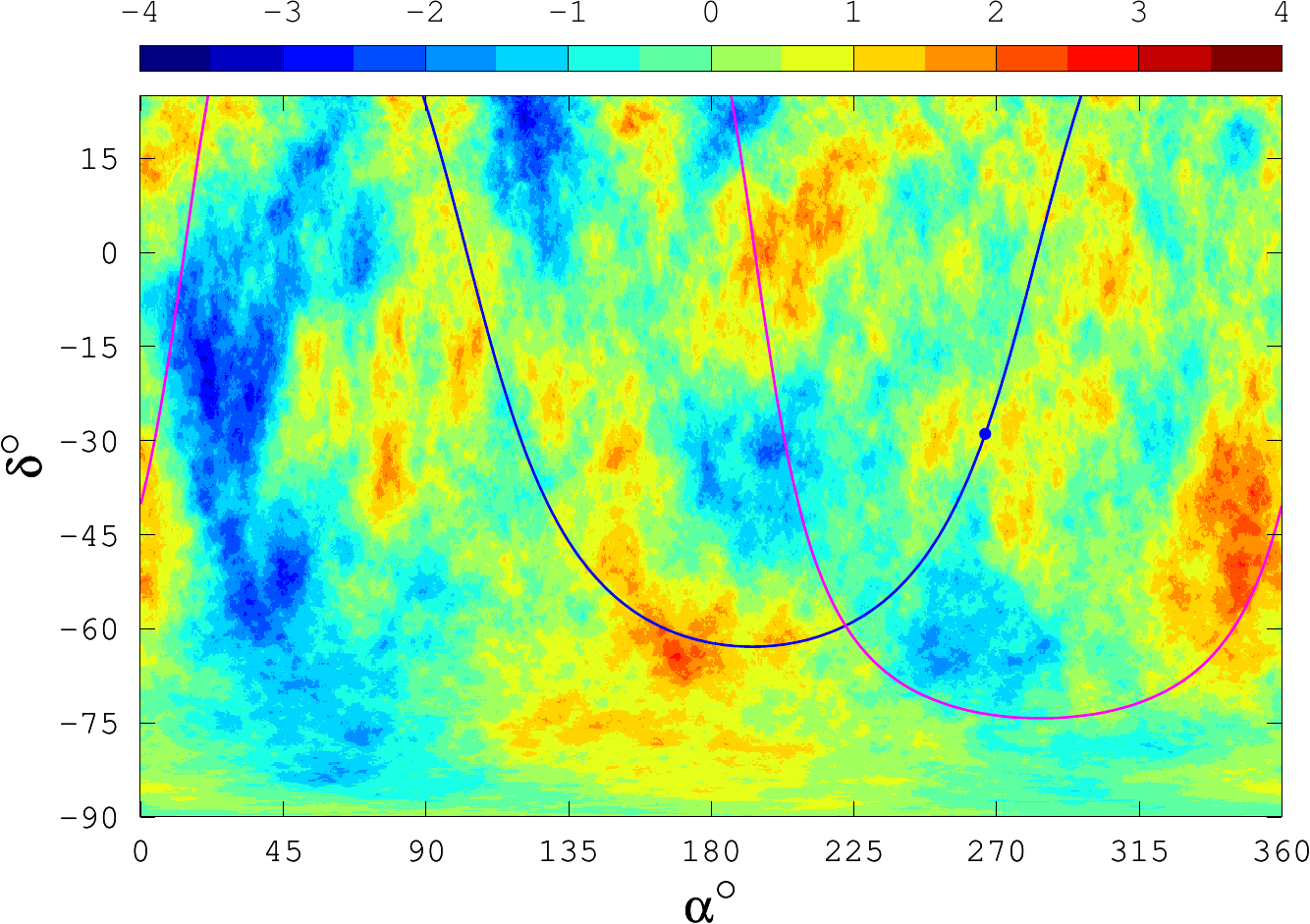}}
	\cl{\fig{.42}{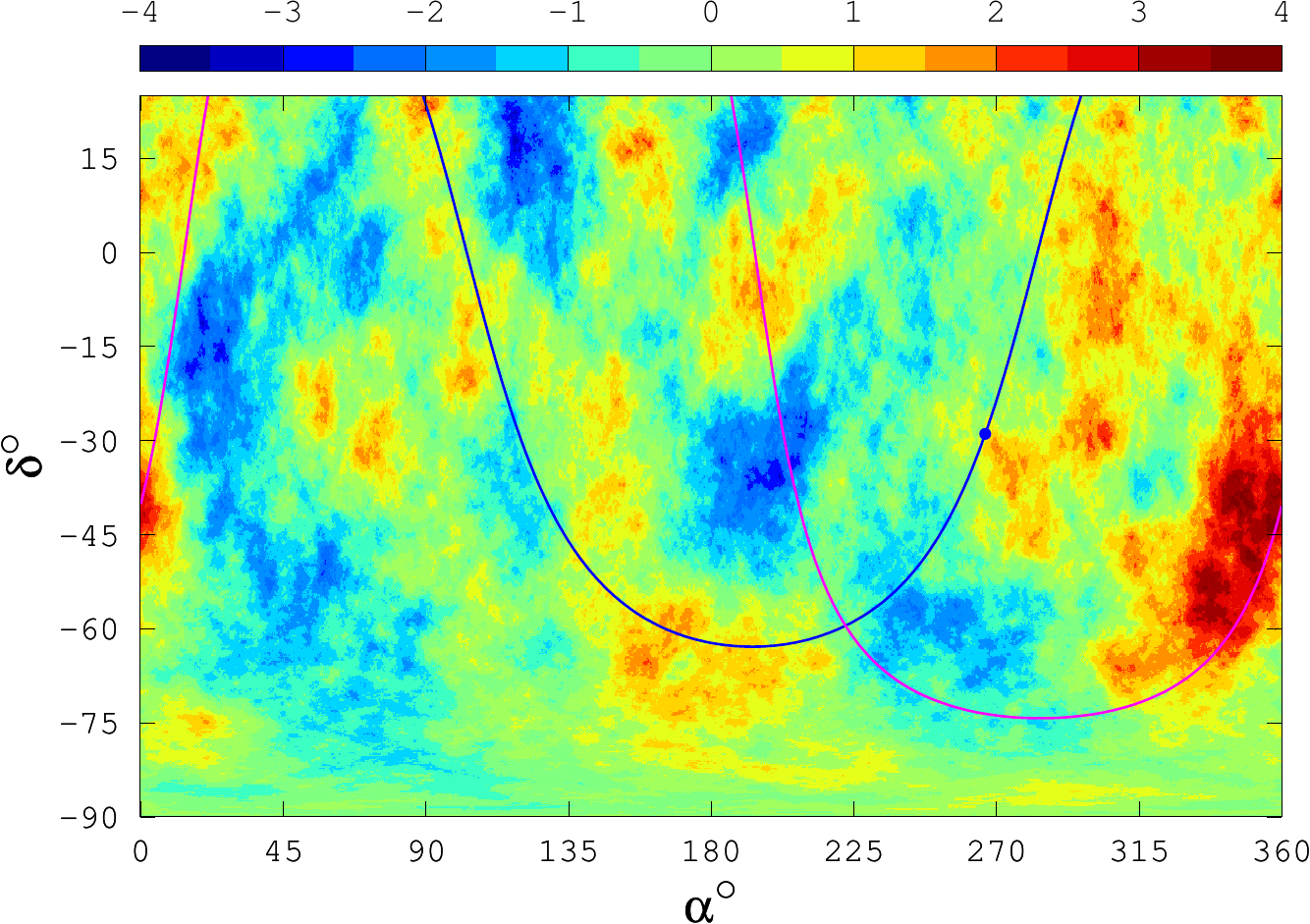}	\fig{.42}{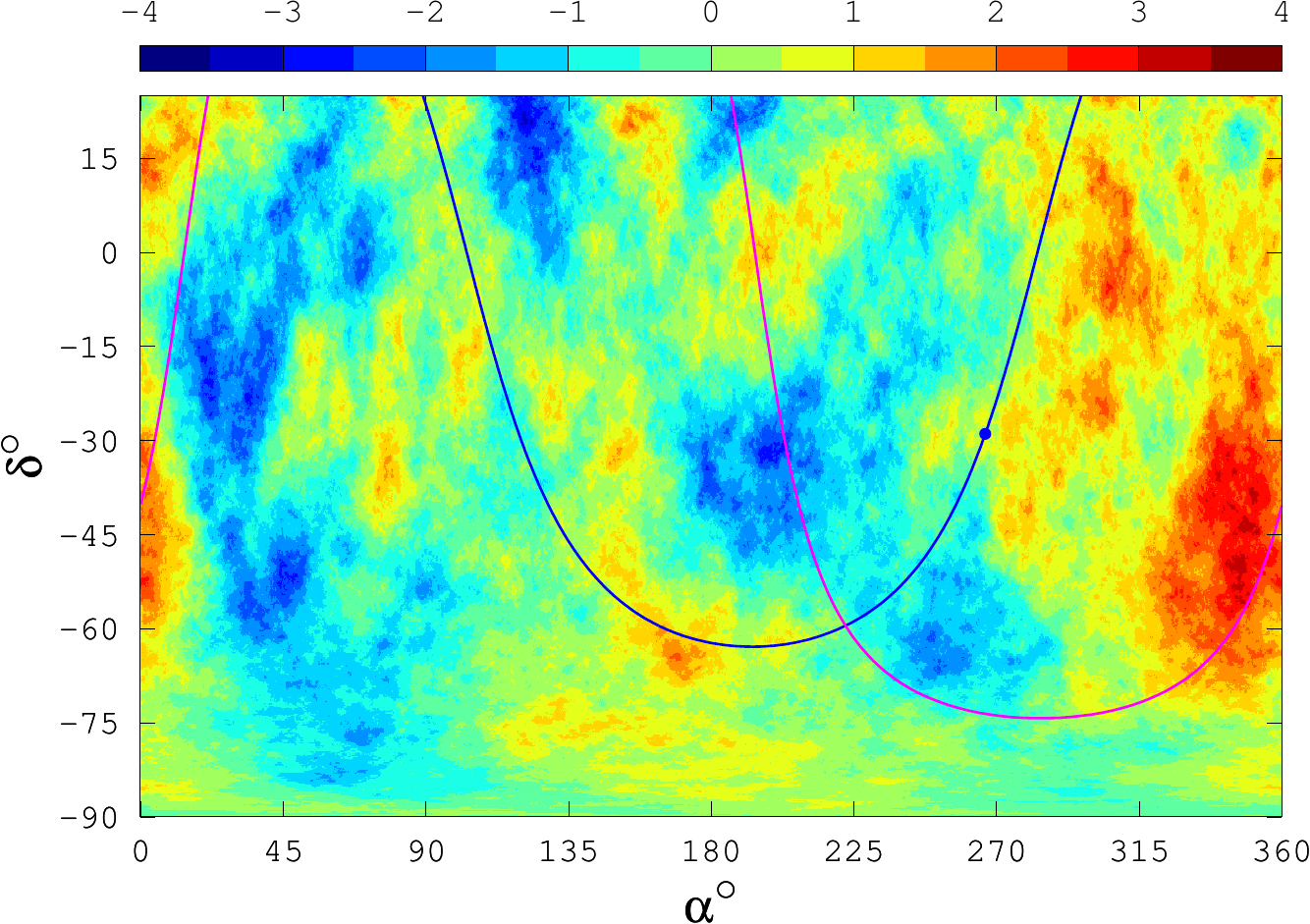}}
	\cl{\fig{.42}{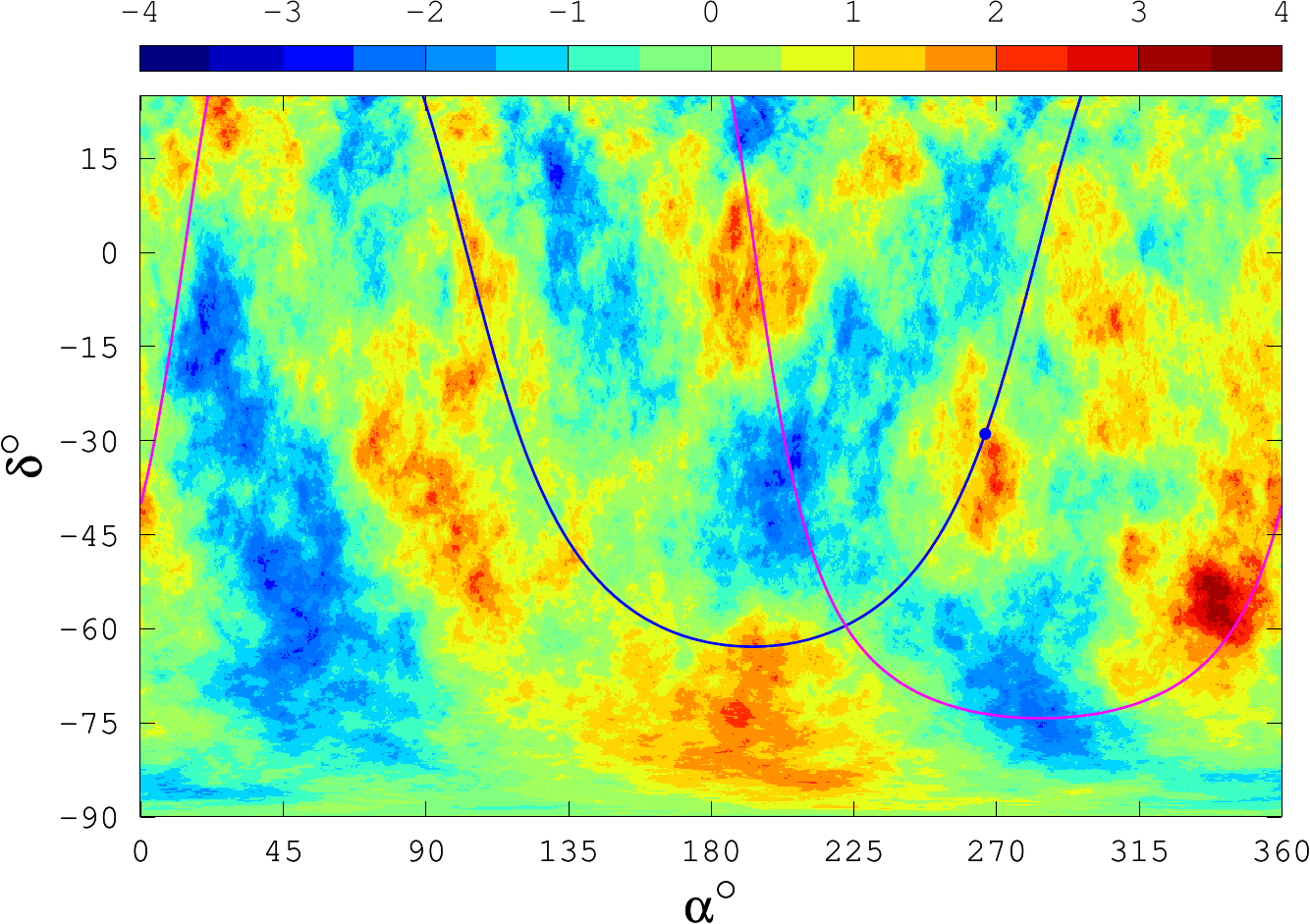}		\fig{.42}{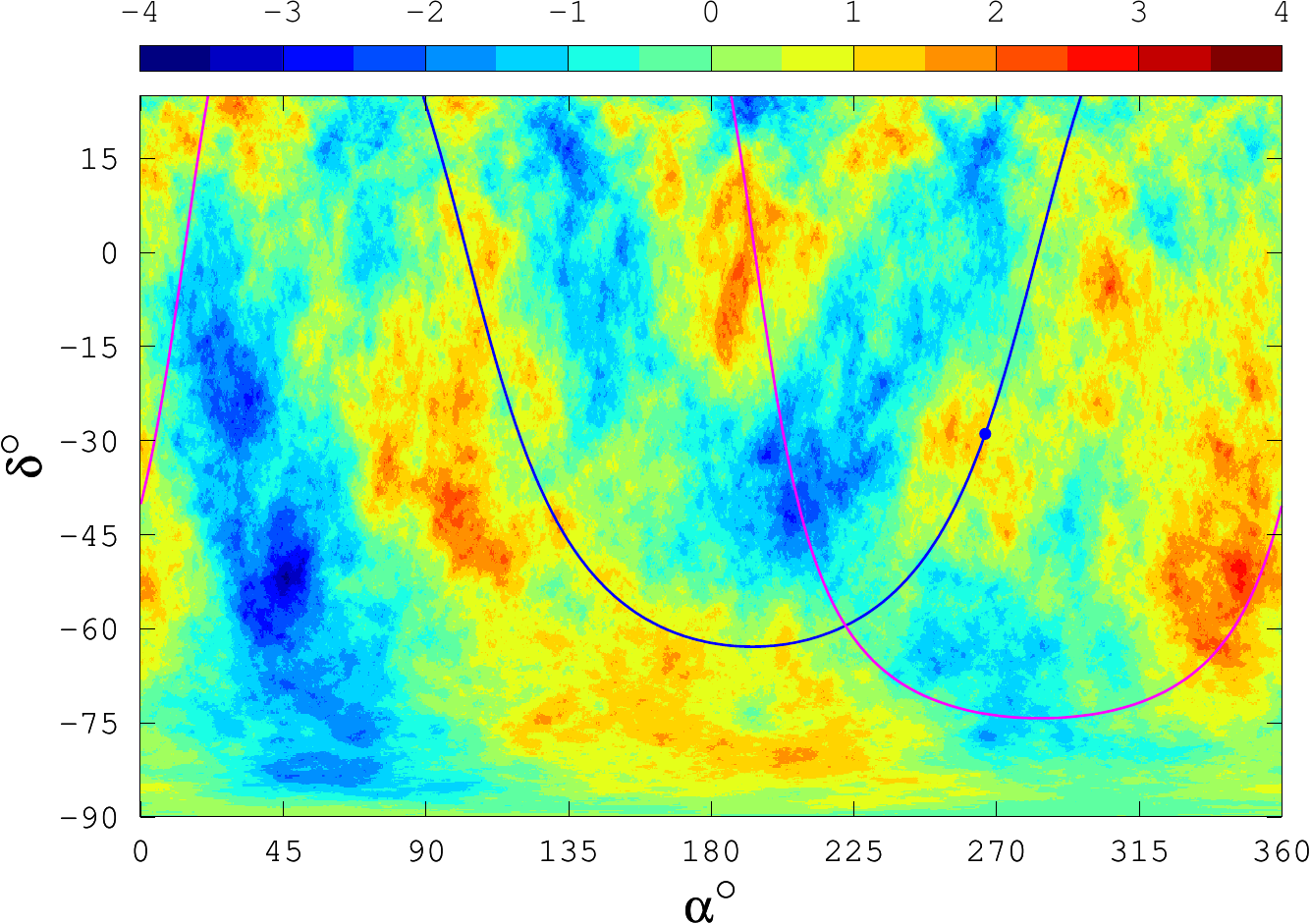}}
	\cl{\fig{.42}{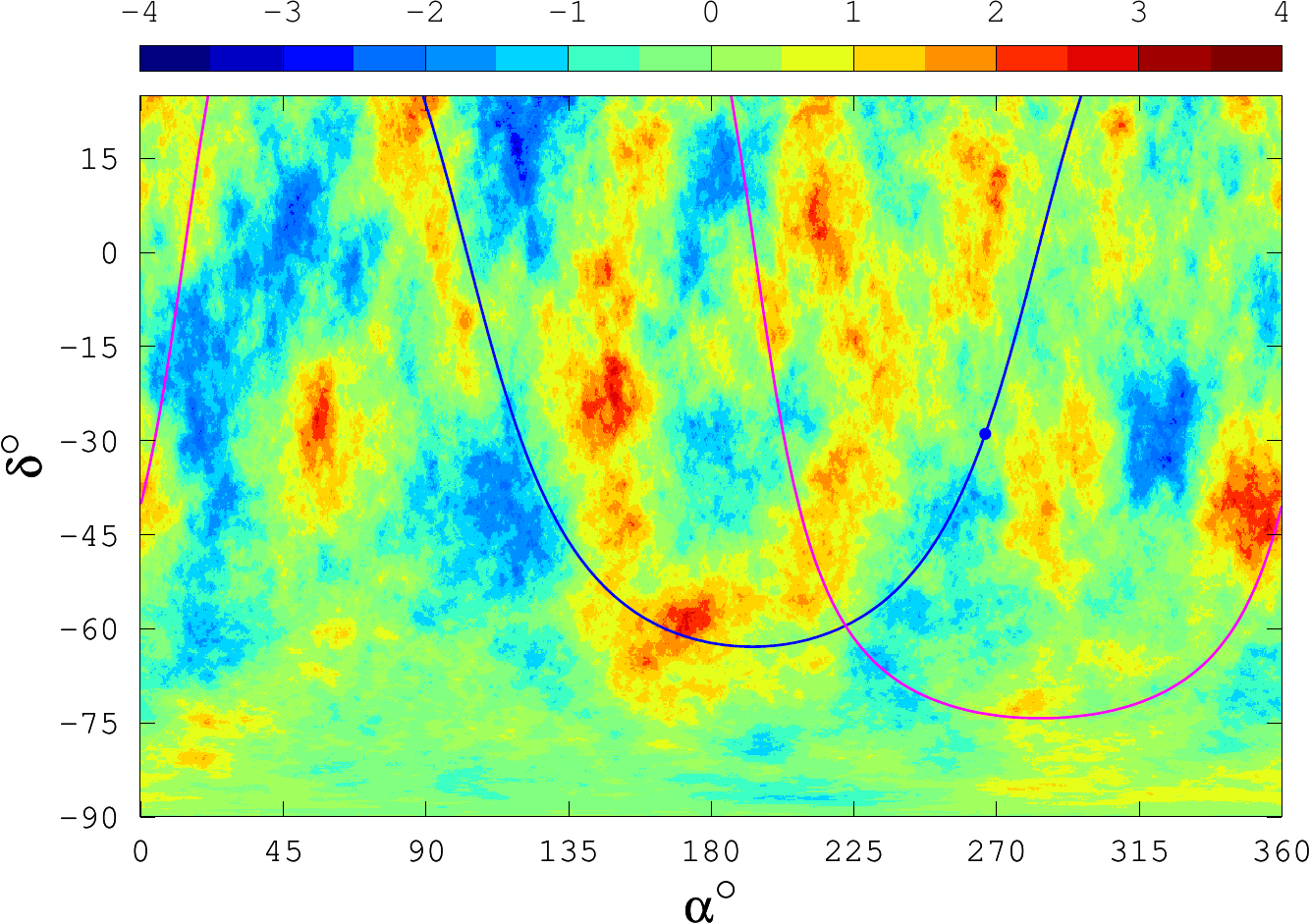}		\fig{.42}{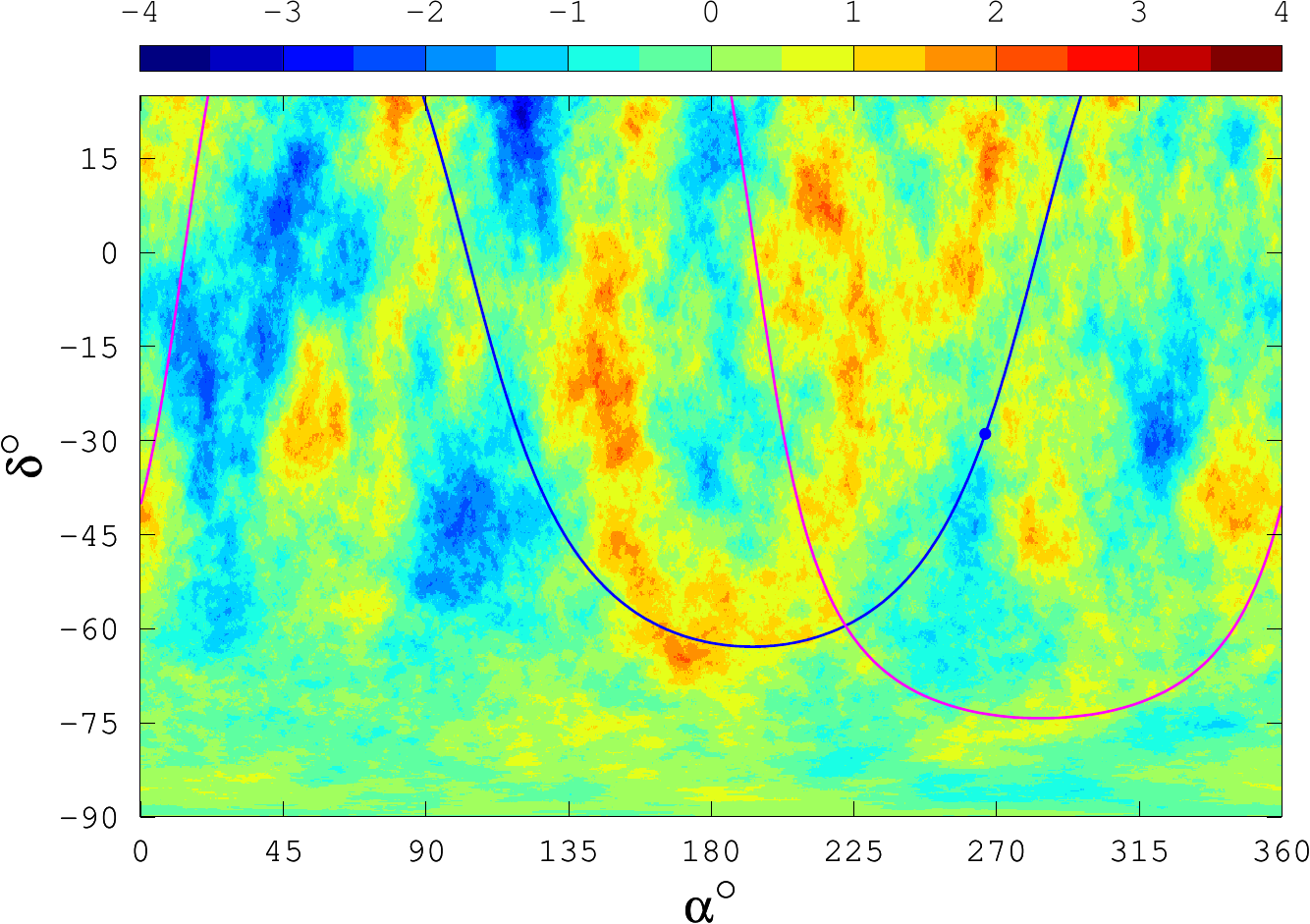}}
\caption{%
	The same as in Fig.~\ref{fig:R8R12} but
	the maps are smoothed at $R=16^\circ$ (left column) and $R=20^\circ$ (right column).
}
\label{fig:R16R20}
\end{figure}

The whole FoV of the Pierre Auger Observatory was studied with circular windows
of different radii~$R$.
Figure~\ref{fig:R8R12} shows anisotropy found for $R=8^\circ$ and $R=12^\circ$.
Figure~\ref{fig:R16R20} shows the same for $R=16^\circ$ and $R=20^\circ$.
The conditions on~$\Nbg$ are satisfied for $\delta\lesssim-6^\circ$,
$\lesssim10^\circ$, $\lesssim15^\circ$ and $\lesssim20^\circ$ for the four
above radii respectively.

\begin{table}[!ht]
	\caption{The most pronounced parts of ``hot spots'' on the maps smoothed
	with $R=8^\circ$, see Fig.~\ref{fig:R8R12}.
	$\Non$~is the number of real events in a circular window centered on
	$(\alpha,\delta)$.
	$\Nbg$~is the same on the background map.
	}
	\label{tab:R8}
	\begin{center}
	\begin{tabular}{|c|c|r|c|c|c|c|}
		\hline
		Data set & Region & $\alpha$ $(^\circ)$ & $\delta$ ($^\circ$) & $\Non$ & $\Nbg$ & $S$ \\
		\hline
		Main
		&  A  &  64.20  & -28.20  & 321.00  & 262.41   &  3.48\\
		&  B  & 179.00  & -65.40  & 379.00  & 319.28   &  3.23\\
		&  C  & 194.60  & -13.40  & 280.00  & 230.71   &  3.13\\
		&     & 191.40  &  -5.80  & 246.00  & 199.61   &  3.16\\
		&  D  & 347.60  & -48.20  & 389.00  & 319.76   &  3.72\\
		&     & 355.80  & -35.20  & 354.00  & 286.24   &  3.84\\
		\hline
		Main (IBG)
		& A'  &  64.20  & -28.20  & 321.00  & 264.38  &   3.35\\
		& B'  & 179.00  & -65.40  & 379.00  & 323.23  &   3.00\\
		& C'  & 194.60  &  -4.20  & 238.00  & 192.94  &   3.12\\
		& D'  & 347.60  & -48.20  & 389.00  & 313.55  &   4.08\\
		&     & 355.80  & -35.20  & 354.00  & 280.68  &   4.18\\
		\hline
		LE
		& B(LE)  & 175.60  & -73.80  & 173.00  & 133.81  &   3.22\\
      &        & 195.80  & -71.60  & 189.00  & 144.35  &   3.53\\
		& C(LE)  & 200.60  &  -6.00  & 134.00  & 101.19  &   3.09\\
		& D(LE)  & 323.20  & -58.40  & 222.00  & 174.83  &   3.40\\
		\hline
		HE
		& A(HE)  &  64.60  & -28.40  & 171.00  & 127.32   &  3.66\\
		& C(HE)  & 188.00  & -13.80  & 153.00  & 110.98   &  3.75\\
		& D(HE)  & 355.80  & -35.20  & 179.00  & 133.34   &  3.74\\
		& HE(1)  & 146.20  & -32.00  & 166.00  & 128.66   &  3.13\\
		& HE(2)  & 158.00  & -56.60  & 193.00  & 153.41   &  3.06\\
		& HE(3)  & 291.20  & -35.60  & 173.00  & 134.29   &  3.18\\
		\hline
	\end{tabular}
\end{center}
\end{table}


\begin{table}[!ht]
	\caption{The most pronounced parts of ``hot spots'' on the maps smoothed
	with $R=12^\circ$, see Fig.~\ref{fig:R8R12}.}
	\label{tab:R12}
	\begin{center}
	\begin{tabular}{|c|c|r|c|c|c|c|}
		\hline
		Data set & Region & $\alpha$ $(^\circ)$ & $\delta$ ($^\circ$) & $\Non$ & $\Nbg$ & $S$ \\
		\hline
		Main
		& C  & 198.60 &   -7.80  & 535.00  & 460.70  &  3.35\\
		& D  & 352.40 &  -41.20  & 788.00  & 677.17  &  4.10\\
		\hline
		Main (IBG)
		& D' & 352.40 &  -41.20  & 788.00  & 664.18  &  4.61\\
		\hline
		LE
		& B(LE)	& 193.60  & -73.60  & 359.00  & 302.13   &  3.14\\
		&	 		& 186.60  & -68.00  & 410.00  & 347.21   &  3.24\\
		& D(LE)	& 342.20  & -55.40  & 460.00  & 386.42   &  3.58\\
		& G		& 263.60  & -38.60  & 393.00  & 333.42   &  3.14\\
		& 			& 264.40  & -35.60  & 385.00  & 325.40   &  3.17\\
		\hline
		HE
		& D(HE)  & 349.80  & -41.60  & 389.00  & 321.72   &  3.59\\
		& HE(1)  & 145.40  & -25.60  & 335.00  & 281.64   &  3.06\\
		\hline
	\end{tabular}
\end{center}
\end{table}


In the main data set, there are four extended regions of excess (``hot spots'')
that deviate from the background by more than~3 standard deviations for
$R=8^\circ$, with Region~D being the most extended and pronounced of them, 
see the top left panel in Fig.~\ref{fig:R8R12} and Table~\ref{tab:R8}.
Its two ``peaks'' are separated from each other by $\approx14^\circ$.
Two of the hot spots, namely Region~C and Region~D, are 
located near the Supergalactic plane (SGP).
An extended Region~B can be seen near the GP.
Region~A is located far from both GP and SGP.

The map of anisotropy obtained for the main data set with the IBG is close
to that obtained with the shuffling technique (see the second row in
Fig.~\ref{fig:R8R12}) but Region~D becomes more pronounced while the Li--Ma
significance decreases for the other hot spots.

There is no counterpart of Region~A in the LE data set but Region~B becomes more
extended and pronounced.
To the contrary, both hot spots located near the SGP become less noticeable.
Region~D splits into two parts with the most pronounced of them being shifted
to higher declinations along the SGP.
The picture of anisotropy for the HE set is considerably different.
Regions~A and~C become much more pronounced than in the main data set while
Region~B practically ``dissolves.''
Region~D becomes more compact with its ``hottest'' part coinciding with one of
the peaks in the main data set.
The HE set has three other hot spots but~$S$ only marginally exceeds
three standard deviations for them, see Table~\ref{tab:R8}.

\clearpage

One might expect that hot spots become less pronounced as the angular scale of
windows used for the analysis grows but this does not happen to be exactly the
case.
While the Li--Ma significance for Region~A and Region~B of the main data set
decreases well below three standard deviations
for $R=12^\circ$, Regions~C and~D become more pronounced, see the top right
panel in Fig.~\ref{fig:R8R12} and Table~\ref{tab:R12}.
For the main data set on the IBG, only Region~D still has $S>3$ and its
deviation from the background is also higher than for $R=8^\circ$.
Notice that Regions~B and~D are still present in the LE data set but
only Region~D and Region~HE(1) remain in the HE set.
One more hot spot appears in the LE set for $R=12^\circ$, see Table~\ref{tab:R12}.
It is located in the GP with its most pronounced part being
$7^\circ$--$10^\circ$ from the direction to the Galactic Center
(on the opposite side of the GP w.r.t.\ the GC comparing with what
was found by AGASA and SUGAR~\cite{AGASA-1999,SUGAR-2001} but not confirmed
later by Auger~\cite{Auger-galcenter-2005,Auger-galcenter-2007}).
The region can also be seen for $R=8^\circ$ both in the LE and in the main
data sets but its deviation from the background is less than three standard
deviations.
Conversely, the flux does not deviate from the background at this location
in the HE data set.

Figure~\ref{fig:R16R20} shows maps of anisotropy obtained for circular windows
of radii $R=16^\circ$ and $R=20^\circ$.
It can be seen that the picture becomes closer to what is expected for isotropy at
large scales as~$R$ grows.
(The Li--Ma significance can be slightly underestimated
since the number of events in regions of these radii becomes comparatively
large, see a remark above.)
There remain no regions in the HE set that deviate from the background by more
than~$3\sigma$.
Only Region~D remains in the other data sets for $R=16^\circ$ with the Li--Ma
significance decreasing.
For $R=20^\circ$, $S<3$ for Region~D in the main data set (on the shuffled
background) and only marginally exceeds~3 for the LE set.

A recent study of the large-scale distribution of CRs
around 1~EeV performed by the Pierre Auger Collaboration
revealed that while it does not contradict isotropy, declination
and right ascension of the dipole for the 1--2~EeV energy band are
$\alpha\approx340^\circ$, $\delta\approx-35^\circ$ with uncertainty
$\sim12^\circ$~\cite{Auger-deAlmeida-2013}.
The direction correlates with Region~D and its counterpart in the HE set.
This allows one to suggest that Region~D is somehow related to the dipole
anisotropy in this energy range.

One can see by comparing Fig.~\ref{fig:R8R12} and Fig.~\ref{fig:R16R20} how
some of the separate regions with an excess or deficit of cosmic rays
concatenate as the angular scale of the analysis increases.
It is especially clear from Fig.~\ref{fig:R16R20} how different are the
``patterns'' of anisotropy for the LE and HE data sets even though they
do not significantly deviate from isotropy.

\section{Conclusions}

The presented results demonstrate that anisotropy of cosmic rays with energies
around 1~EeV might have interesting features at certain intermediate angular
scales, including localized regions of excess, and that anisotropy evolves
with energy.
Still, the study has a serious limitation because of the small amount of data used.
On the one hand,
calculations performed with other available data reveal that a sample
consisting of just one or a few percent of the whole data set only partially
reproduces anisotropy of the whole set, and significant differences
are possible.
On the other hand, such limited data sets as those studied do not allow one to apply
more advanced methods of analysis.
In particular, it is hardly possible to accurately estimate the dipole anisotropy
and to take it into account when studying anisotropy at intermediate scales.
In our opinion, it would be interesting if an analysis of anisotropy of cosmic
rays around 1~EeV at intermediate angular scales is performed with full data
sets of the Auger and Telescope Array experiments.
It can provide information concerning the structure of the galactic magnetic
field and the fundamental problem of the transition from galactic to
extragalactic cosmic rays.

\bigskip

I am glad to thank Bohdan Hnatyk and Peter~Tinyakov for stimulating discussions,
Miguel Mostaf\'a for help with the data at the preliminary stage of the work,
and Daniel~Fiorino for the useful communication concerning presentation of the
results.
GNU~Octave was used for the calculations~\cite{Octave}.

The study was made with a partial financial support
by the Russian Foundation for Basic Research grant 13-02-12175 ofi\_m.


\end{document}